\documentclass[journal=jacsat,manuscript=article
]{achemso}

\usepackage{chemformula} % Formula subscripts using \ch{}
\usepackage[T1]{fontenc} % Use modern font encodings

\author{Silvana A. Caipa Cure}
\affiliation[Leiden University]
{Department of Soft and Biological Matter, Huygens-Kamerlingh Onnes Laboratory, Leiden University, PO Box 9504, 2300 RA Leiden, the Netherlands}
\author{Daniela J. Kraft}
\affiliation [Leiden University]
{Department of Soft and Biological Matter, Huygens-Kamerlingh Onnes Laboratory, PO Box 9504, 2300 RA Leiden, the Netherlands}
\email{Kraft@physics.leidenuniv.nl}

\title[An \textsf{achemso} demo]
  {Fabrication and characterization of bimetallic silica-based and 3D-printed active colloidal cubes}

\begin{document}

\begin{abstract}
 Simulations on self-propelling active cubes reveal interesting behaviors at both the individual and the collective level, emphasizing the importance of developing experimental analogs that allow to test these theoretical predictions. The majority of experimental realizations of active colloidal cubes rely on light actuation and/or magnetic fields to have a persistent active mechanism, and lack material versatility. Here we propose a system of active bimetallic cubes whose propulsion mechanism is based on a catalytic reaction and study their behavior. We realize such a system from synthetic silica cuboids and 3D printed micro-cubes, followed by the deposition of gold and platinum layers on their surface. We characterize the colloids' dynamics for different thicknesses of the gold layer at low and high hydrogen peroxide concentrations. We show that the thickness of the base gold layer has only a minor effect on the self-propulsion speed and in addition induces a gravitational torque which leads to particles with a velocity director pointing out of the plane thus effectively suppressing propulsion. We find that the a higher active force can remedy the effects of torque, resulting in particle orientations that are favorable for in plane propulsion. Finally, we use 3D printing to compare our results to cubes made from a different material, size and roundness, and  demonstrate that the speed scaling with increasing particle size originates from the size-dependent drag. 
 Our experiments extend fabrication of active cubes to different materials and propulsion mechanisms and highlight that the design of active particles with anisotropic shapes requires consideration of the interplay between the shape and activity to achieve favorable sedimentation and efficient in-plane propulsion. 
\end{abstract}

\section{Introduction}
 
In recent years, much effort has been put into the fabrication and characterization of synthetic micro-swimmers that are able to move autonomously and in a controlled fashion\cite{Buttinoni2012, Palacci2013_livingcrystals, theurkauff_dynamic_2012}. These active systems are appealing not only for their out-of-equilibrium dynamics \cite{Bechinger2016,theurkauff_dynamic_2012,gonnella_phase_2014} but also on account of their ability to model biological and physical systems \cite{gompper_2020_2020,Ramaswamy2010}, biomedical cargo transportation\cite{Xu2019} and environmental remediation techniques\cite{Katuri2016}. Experimentally, a range of approaches have been explored to propel artificial micro-swimmers: from colloids that are activated by light or magnetic fields \cite{Dreyfus2005,Zhang2019}, to swimmers with built-in polarity such as Janus particles that are capable of exhibiting self electro-, thermo-, diffusio- and chemo-phoresis\cite{Howse2007,Paxton2004}.\\

A significant fraction of experimental studies on self-propelled particles have focused on colloidal spheres and rods: from the early work of \textit{Howse et al.} on Janus spherical particles with a catalytic platinum (Pt) patch \cite{Howse2007}, to the development of rod shape particles consisting of gold (Au) and platinum segments\cite{Paxton2004}, to more cutting edge fabrication techniques such as mesoporous silica spheres where the Pt is situated on the inside\cite{Ma2016}. In contrast, experimental research on other geometries is quite limited, although progress has been made with fabricating active anisotropic particles, for example prepared by assembling spheres\cite{Wang2019,Alvarez2021} or based on litography and 3D micro-printing as it is the case for L-shaped\cite{Kmmel2013}, helical\cite{Doherty2020}, tori\cite{Wang2022} and crescent-shaped designs\cite{Riedel2024}.\\

Active colloidal cubes exhibit a rather simple form of shape anisotropy, yet a variety of interesting emergent properties have been predicted for these microswimmers. 3D simulations of active sphero-cubes in confinement showed unusual phase behavior with a second-order freezing transition, and predicted new crystal structures such as the sheared cubic crystal at high densities\cite{Wensink2013}. Moreover, simulations of hard active cubes in 2D have yielded compelling findings on the influence of shape on collision efficiency, which leads to lower critical packing fractions than those observed for active hard spheres for the onset of mobility induced phase separation (MIPS), and the formation of multiple stable clusters during MIPS that persist for long timescales\cite{Moran2022Glotzer}. Therefore, experimental model systems that allow to test these theoretical predictions are considered highly valuable.\\

Experimentally, haematite ($\alpha$-Fe\textsubscript{2}O\textsubscript{3})-based colloidal particles are by far the most successful realization of self-propelled colloidal cubes. Although not truly cubic in shape, composite colloids made from haematite cubes encapsulated by a polymer sphere made from 3-methacryloxypropyl trimethoxysilane (TPM) were shown to  exhibit self propulsion when exposed to a UV light source in a hydrogen peroxide (H\textsubscript{2}O\textsubscript{2}) medium.\cite{Palacci2013_livingcrystals,Palacci2014} Research that followed showed that their active motion can be controlled by taking advantage of the permanent magnetic moment of this iron oxide to perform tasks such as cargo docking and transport\cite{Palacci2013}. A more recent experimental realization of this system showed that the deposition of Pt metal on one of the sides of the haematite cubes enhances their activity in the presence of chemical fuel and when illuminated by a UV source\cite{Zhang2023}. Despite these advancements on haematite cubic colloids there is no realization of active cubes at the microscale that (i) does not depend on light activation to have a persistent active mechanism and (ii) consists of materials other than haematite. Light-independent persistent self-propulsion is key to probe particles' behavior within setups in which having a UV source is not possible or not desired due to photo-bleaching of fluorescent dyes, sample heating or observation of light sensitive-reactions. Additionally moving away from haematite into more versatile materials expands the fabrication possibilities for this kind of anisotropic particles.\\

In this paper we combine experimental techniques to fabricate colloidal cubes with electro-catalytic self propulsion mechanisms to introduce activity. For this we use both synthetic silica cuboids templated from haematite and 3D printed micro-cubes to achieve a cubic shape, followed by the deposition of gold and platinum on the cubes' surface to implement the active mechanism. 
%The resulting colloids have both self diffusio- and electro-phoretic propulsion. Briefly, the platinum side is employed to catalyze the decomposition of hydrogen peroxide and the gold side is used as an anodic end, generating electrochemical gradients that promote the phoretic movement of the particle\cite{Moran2010}. 
We qualitatively describe and quantify the system's dynamics for different thicknesses of the Au layer at low and high hydrogen peroxide concentrations. Our observations show that, for a low hydrogen peroxide concentration, cuboid colloids do not undergo orientational or dynamical changes even for increasing thickness of the metallic layer on their surface, yielding similar propulsion speeds. However as the hydrogen peroxide concentration is increased, the particles' speeds follow a bi or tri-nodal distribution hinting at a different sedimentation behavior. We also study the effect of material, size and shape on the cubes' self-propulsion speeds, by comparing the synthetic particles to 3D-printed ones and demonstrate that decreasing speeds with increasing particle size originate from an increased drag. Our experimental technique provides an alternative route to fabricate active cubes from versatile materials such as silica and polymers, expanding the fabrication possibilities beyond those currently available, removing the need for light activation, and allowing for experimental validation of theoretical predictions at high particle densities.
 
\section{Experimental details}
\textbf{Materials.} Hydrogen peroxide (H\textsubscript{2}O\textsubscript{2}, $35\%$, in water), and propylene glycol monomethylether acrylate (PGMEA, >$99.5\%$) were purchased from Sigma-Aldrich. 2-propanol (IPA,$9,99\%$) was purchased from VWR Chemicals. All materials were used as received unless stated otherwise. All solutions were prepared from deionized water with 18.2M$\Omega$ cm resistivity, using a Millipore Filtration System (Milli-Q Gradient A10).

\hspace{0.5 in}

\noindent\textbf{Preparation of Silica Cuboids}. Colloidal cuboids with edge-to-edge length 2.0 \textpm 0.3 \textmu m were previously synthesized for the work of \textit{Shelke et al.} \cite{doi:10.1021/acsnano.3c00751}. Briefly, mono-disperse pseudo-cubic haematite particles (edge-to-edge length 1.52 \textpm 0.05 \textmu m) were prepared from condensed ferric hydroxide gel \cite{Sugimoto1992}, and then coated by a silica layer by a St\"ober procedure \cite{Wang2013}. After dissolution of the haematite cores with HCl, hollow silica shells with cuboid geometry were obtained. The shape of these cuboids lies in between a sphere and a cube and can be described by $(2\frac{x}{L})^m + (2\frac{y}{L})^m + (2\frac{z}{L})^m \leq 1$, were $L$ is the face-to-face length and $m$ is a shape parameter typically ranging between $2.5<m<4$\cite{Meijer2019} and $L=2.0 \pm 0.3 \mu m$.

\hspace{0.5 in}

\noindent\textbf{Particle printing}. Cubes with sharp edges and corners were fabricated using a commercially available two-photon polymerization (2PP) 3D micro-printer (Photonic Professional Gt, Nanoscribe GmbH) equipped with a 63x oil-immersion objective (Zeiss, NA=1.4) in dip mode. Conditional to the resolution of the equipment the cubes were printed with a side length of 4 \textmu m. First, the cubes were designed and rendered using Autodesk Inventor and Describe. Then, the structures were printed onto a fused silica substrate using the commercial photo-resist IP-Dip purchased from Nanoscribe GmbH\cite{Doherty2020}. After printing, the particles were developed by 30 min submersion in PGMEA and a 2 min dip into IPA, and subsequently left to dry. Once developed, the particles were ready for sputter coating. 

\hspace{0.5 in}

\noindent\textbf{Metallic layer deposition}. Cuboid silica particles and 3D printed particles were sputter-coated with thin metallic layers to integrate a propulsion mechanism. For this, 50 $\mu$L of diluted colloidal dispersion were deposited onto a silica substrate (2.5 cm x 2.5 cm) via spin-coating using a SCS 6800 Spin coater series at 1983 RPM for 30 s. The concentration was tuned manually such that spincoating resulted in a mono-layer. This step was not necessary for the 3D printed particles because they were already distributed in a monolayer on top of a silica substrate.
\\
The substrate containing the monolayer  of particles was then sputter-coated  using the following vacuum systems consecutively. 
We used a high vacuum Leybold-Heraeus Z400 sputter-coater with custom modifications to obtain gold coating of all exposed surfaces of the cubes. In this system the target is close to the substrate and sputtering is performed at $\approx 10^{-5}$ mbar (Ar @ 54 sccm, Direct Current Electrde Positive (DCP): 1kV) yielding a robust layer that covers most of the particle. This results in cuboids that have five out of six faces covered by a gold layer.
We secondly employed a low vacuum ($10^{-2}$ mbar) Cressington 208HR sputter-coater to  create a thin metallic cap of platinum for coating only the top of the particles. The Pt target was purchased at Micro to Nano (\O 57 x 0.2 mm, $99.99\%$). 
The consecutive sputtering with gold and platinum in the two systems yielded particles that had five out of six faces covered by a gold layer, and on the top half a platinum layer. Different metal layer thicknesses were tested: 20 nm - 500 nm of Au (in high vacuum)  and 5 nm - 20 nm of Pt (in low vacuum). A schematic of this procedure can be found in Figure \ref{fig:fig2}A.
\\ 
Sputter-coated particles were then recovered from the substrate via sonication and re-dispersed in water.

\hspace{0.5 in}

\noindent\textbf{Particle observation.} To characterize the surface of the particles after sputter-coating, scanning Electron Microscopy (SEM) images were taken with a Thermo-Fisher Apreo SEM. For this, particles were deposited onto a silicon SEM substrate and allowed to dry in air before imaging.
\\
Information on the dynamics of the metal-coated particles was extracted from brightfield microscopy observations in a custom made microscope holder in the presence of 1\%v/v or 5\%v/v hydrogen peroxide   (H\textsubscript{2}O\textsubscript{2}) in water. A fresh H\textsubscript{2}O\textsubscript{2} solution was made for every measurement day, to avoid depletion of the fuel due to light or temperature. Particles were imaged using a Nikon Eclipse Ti microscope with 60x water immersion (NA=0.7) and 20x dry (NA=0.5) objectives.  Movies of 30 s to 120 s were taken for all samples at 20 frames per second (fps), unless stated otherwise. The surface area fraction of particles was $\phi_{particles}=0.01\%-0.05\%$ in relation with the total area of the field of view. Samples were only kept for a maximum of 30 min to avoid H\textsubscript{2}O\textsubscript{2} depletion, convection effects due to bubble formation, and because an increasing number of particles became stuck to the substrate. 

\hspace{0.5 in}

\noindent\textbf{Data analysis} The videos were analyzed using the Crocker-Grier algorithm implemented for Python as Trackpy\cite{https://doi.org/10.5281/zenodo.7670439} to extract XY trajectories of the particles. Although Trackpy is intended to track particles with round features, at the magnifications used, the features could be approximated to circles yielding reliable tracking results. From the trajectories identified with the algorithm, mean square displacements (MSD) were calculated and fitted to $\langle r^2\rangle= 4D_{0}\delta t+ v^{2}\delta t^{2}$, where $D_{0}$ is the translational diffusion coefficient, $v$ is the particle speed and $\delta t_{max}= 1s $ is the maximum lag time for the fit, which is lower than the rotational diffusion time of the particles\cite{Howse2007, Bechinger2016}. The fit allowed us to estimate the particles' speed ($v$) and their translational diffusion coefficient ($D_{0}$). For all samples, >30 particles were analyzed per sample to get statistically relevant information, with the exception of printed particles for which sometimes it was possible to obtain only $n\approx10$, given the low particle density that the 3D micro-printing methods yield. Particles that were stuck to the substrate, i.e not displaying activity or Brownian motion, were not taken into account during the analysis.

\hspace{0.5 in}

\section{Results and Discussion}

\begin{figure}[H]
    \centering
    \includegraphics[scale=1]{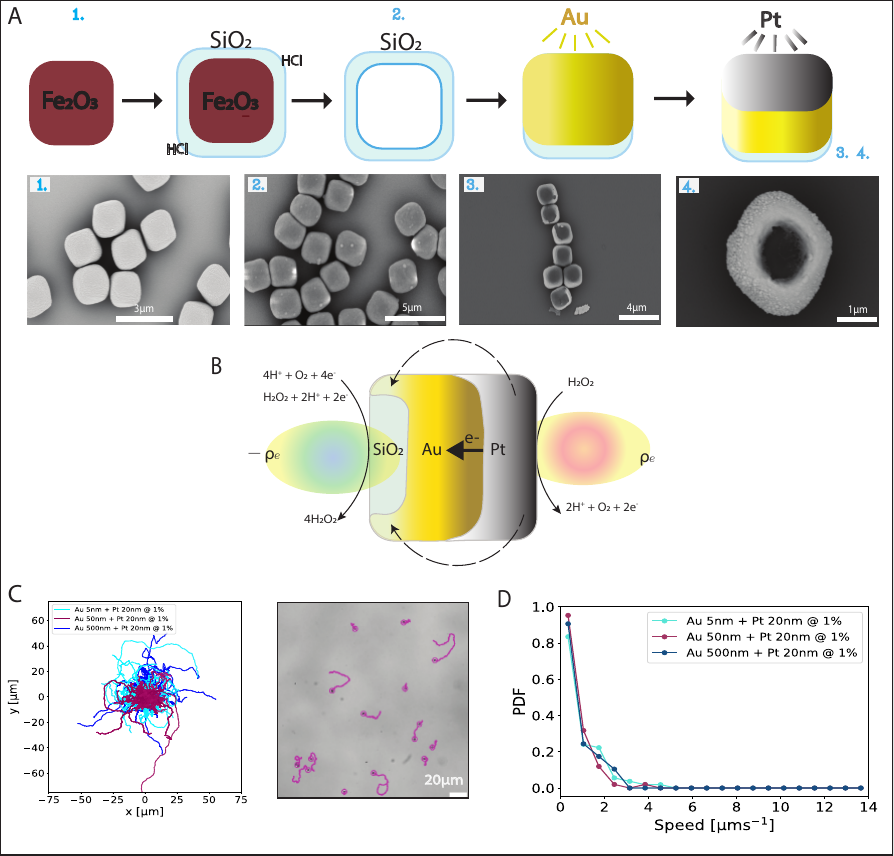}
    \caption{(A) Schematic of the fabrication process for silica- based bimetallic cuboids, accompanied by SEM images of (A.1) haematite cuboids (A.2) haematite cuboids coated with silica and hollow silica cuboids after acid dissolution of the haematite core (A.3) metal coated silica cuboids with a 5 nm Au base layer followed by a 20 nm Pt top layer and (A.4) a 500 nm Au base layer followed by a 20 nm Pt top layer. (B) Schematic of self-electrophoresis of a bimetallic Au-Pt silica cuboid adapted from Moran, 2017\cite{moran_phoretic_2017}. The hydrogen peroxide is oxidized on the platinum's (cathode) surface driving an electron flow into the gold (anode) layer and a proton flow on the fluid. This reaction leads to an asymmetric charge density distribution that generates electric fields and a slip flow of the fluid around the particle that propels the particle in the opposite direction. (C) Trajectories centered at the origin, recorded over the course of 30 s at 20 fps for active cuboids coated with different thicknesses of an Au layer followed by a 20 nm Pt layer, in the presence of H\textsubscript{2}O\textsubscript{2} 1\%v/v. Accompanied by a sample snapshot of 30 s trajectories depicted on top of the corresponding brightfield image. (D) Speed distributions of >30 particles for each of the layers' thicknesses, in a H\textsubscript{2}O\textsubscript{2} 1\%v/v solution. The speeds were fitted after calculating the MSDs from trajectories recorded via optical microscopy, and fitting them to $MSD= 4D_{0}\delta t+ 2v^{2}\delta t^{2}$  with $\delta t_{max}=$ 1s.}
     \label{fig:fig2}
\end{figure}

\textbf{Dynamics of bimetallic Au-Pt silica-based cuboids with increasing metallic layer thickness}. Our strategy for the fabrication of active cubes consisted of using a combination of two metals on the surface of cuboid silica particles to promote both self diffusio- and electro-phoretic propulsion in a hydrogen peroxide solution. For this, haematite colloids - Figure \ref{fig:fig2} A.1  - were used to template silica cuboids with edge-to-edge length $2.0\pm 0.3$ \textmu m - Figure \ref{fig:fig2} A.2 - via a St\"ober procedure. The cuboids were spin coated onto a glass slide and sputter-coated in a high vacuum system from above to cover 5 out of 6 faces of the particle's surface with a 5 nm Au layer. We then employed a low vacuum sputtering system that allowed us to deposit a 20  nm thick Pt layer on the top half of the cuboids, see Figure \ref{fig:fig2} A.3 and A.4. The corresponding SEM images of each step are shown below the schematic. This technique allowed us to introduce an anodic layer (Au) to promote self-electrophoresis, to the well-established self-propulsion mechanism in which a Pt layer drives the asymmetric decomposition of H\textsubscript{2}O\textsubscript{2}\cite{Ibrahim2017}. Self-electrophoretic activity results from the interaction between a self-generated electric field and a charged colloidal surface. For Au-Pt bimetallic swimmers in a H\textsubscript{2}O\textsubscript{2} solution, a proton concentration gradient as a result of Redox reactions, establishes an electrical dipole around the particle that, when coupled with the free charges present in the particle's electrical double layer, induces an electroosmotic slip around the particle causing it to swim \cite{moran_phoretic_2017,MORAN2011,Moran2010}, see Figure \ref{fig:fig2}B. Additionally, at low fuel concentrations, self-electrophoretic swimmers have proven to propel significantly faster than swimmers driven by other active mechanisms\cite{Bechinger2016,moran_phoretic_2017,klongvessa_nonmonotonic_2019,Wheat2010}, which can  be advantageous for studies at higher particle densities, such as was shown by Klongvessa et al. \cite{klongvessa_nonmonotonic_2019}. High fuel efficiency at relatively low fuel concentration avoids disturbance of the experiments by bubble formation while still allowing for high particle velocity.

 In contact with 1\%v/v H\textsubscript{2}O\textsubscript{2}, bi-metallic silica cuboids with 5 nm Pt and 20 nm Au layers displayed self propulsion. Their dynamics were observed by brightfield microscopy and their position was tracked using Trackpy, as shown in Figure \ref{fig:fig2}C - light blue. The speeds are extracted from their short-time mean-squared displacement and a probability density function of the speed was obtained from > 30 particle trajectories as shown in Figure \ref{fig:fig2}D - light blue. The mean particle speed  extracted from the particles' short-time mean-squared displacement (n>30) was 0.94 \textpm 0.85 $\mu m s^{-1}$. The uncertainty corresponds to one standard deviation and shows the broad distribution of the data.\\

We followed by studying the effect of the thickness of the gold layer on the swimming speed of bimetallic Au-Pt silica-based cuboid particles. The reasons for that were two-fold: First, we were inspired by the high speeds exhibited by Pt coated Au spheres reported by \textit{Klongvessa et al.} and \textit{Theurkauff et al.} where particles attained a speed of up to three body lengths per second at low H\textsubscript{2}O\textsubscript{2} concentrations\cite{klongvessa_nonmonotonic_2019,theurkauff_dynamic_2012}. Second, the effect of the platinum's thickness has already been investigated for electrophoretic systems of bimetallic particles were a directly proportional relation between the catalyst's thickness and the particles' speed has been observed\cite{Zhang2023}.\\

For the Au base layer, two other thicknesses were assessed (50 nm and 500 nm) while the top Pt layer was kept at a constant thickness of 20 nm. 
We found that all cuboid particles posses similar speed distributions regardless of the thickness of the metallic base layer as can be observed by the overlap in Figure \ref{fig:fig2}D. In line with this, the arithmetic mean speeds of 0.94 \textpm 0.85 $\mu m s^{-1}$, 0.71 \textpm 0.59 $\mu m s^{-1}$, 0.76 \textpm 0.64 $\mu m s^{-1}$  for 5 nm, 50 nm, 500 nm Au layer thickness respectively agreed within the error. In addition, all mean speeds were less than half a body length per second and thus significantly smaller than those found by \textit{Klongvessa et al.} and \textit{Theurkauff et al.}\cite{klongvessa_nonmonotonic_2019,theurkauff_dynamic_2012} suggesting that thick layers of gold are not able to recreate the fast propulsion speed properties seen in Pt-coated solid Au particles.\cite{klongvessa_nonmonotonic_2019} 
We conclude that that at 1\%v/v H\textsubscript{2}O\textsubscript{2} the thickness of the gold layer does not have an effect on the translational speed of the cuboids. 

 \hspace{0.5 in}

\begin{figure}[H]
    \centering
    \includegraphics[scale=1]{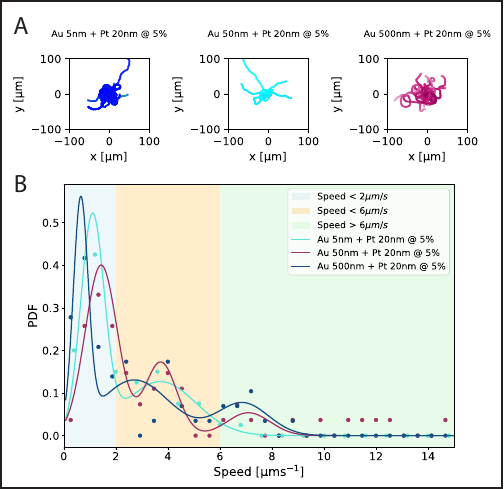}
    \caption{(A) Trajectories centered at the origin, recorded over the course of 30 s at 20 fps for active cuboids coated with a 5 nm, 50 nm and 500 nm Au layer respectively followed by a 20 nm Pt layer, in the presence of H\textsubscript{2}O\textsubscript{2}  5\%v/v. (B) Speed distribution of >30 particles for each of the layer thicknesses configurations, in a H\textsubscript{2}O\textsubscript{2} 5\%v/v solution. For each gold layer thickness we performed a two or three component Gaussian fit as follows $f(x) = A_1 e^{-\frac{(x - \mu_1)^2}{2 \sigma_1^2}} + A_2 e^{-\frac{(x - \mu_2)^2}{2 \sigma_2^2}} + A_3 e^{-\frac{(x - \mu_3)^2}{2 \sigma_3^2}}$. Where $A_1,A_2,A_3$ are the independent  amplitudes, $\mu_1,\mu_2,\mu_3$ the independent means and $\sigma_1,\sigma_2,\sigma_3$ the independent standard deviations. The fit resulted in two peaks for an Au thickness of 5 nm and three peaks for Au thicknesses of 50 nm and 500 nm. The blue background highlights the peaks under 2 $\mu m s^{-1}$, the orange one peaks in between 2 $\mu m s^{-1}$ and 6  $\mu m s^{-1}$, and the green background peaks above 6$\mu m s^{-1}$.}
     \label{fig:fig3}
\end{figure}

\textbf{Dynamics of bimetallic Au-Pt silica-based cuboids with increasing metallic layer thickness at a higher fuel concentration.} 
With the aim of making our system more active, i.e. higher mean speeds, and gaining a better understanding of the active mechanism, we performed the same measurements as for the previous section, but this time for 5\%v/v instead of 1\%v/v H\textsubscript{2}O\textsubscript{2}, as shown in Figure \ref{fig:fig3}. Increasing the fuel concentration up to 10\%v/v in H\textsubscript{2}O\textsubscript{2}-Pt catalytic systems is a reliable way of obtaining higher colloidal speeds \cite{Howse2007,Moran2010}.

Figure \ref{fig:fig3}B depicts the measured speed distributions for the colloidal systems probed at 5\%v/v H\textsubscript{2}O\textsubscript{2}, as extracted by quantitative particle tracking from bright-field microscopy videos. The distributions are obtained from >30 particle trajectories' and the velocity is extracted from their short-time mean-squared displacement. Compared to the data for 1\%v/v H\textsubscript{2}O\textsubscript{2}, the mean speeds of the particles were significantly higher, i.e. 2.24 \textpm 1.61 $\mu m s^{-1}$, 4.00 \textpm 3.88 $\mu m s^{-1}$, 2.75 \textpm 2.42 $\mu m s^{-1}$  for a 5 nm, 50 nm, 500 nm thick Au layer respectively, which represent an average speed increase of 3.8$\times$ upon fuel addition. Again, the uncertainty corresponds to one standard deviation and shows the broad distribution of the data. This time the resulting speed distributions were fitted to a two (5 nm Au layer) or three (50 nm and 500 nm Au layer) component Gaussian fit as follows $f(x) = A_1 e^{-\frac{(x - \mu_1)^2}{2 \sigma_1^2}} + A_2 e^{-\frac{(x - \mu_2)^2}{2 \sigma_2^2}} + A_3 e^{-\frac{(x - \mu_3)^2}{2 \sigma_3^2}}$. Where $A_1,A_2,A_3$ are the independent  amplitudes, $\mu_1,\mu_2,\mu_3$ the independent means and $\sigma_1,\sigma_2,\sigma_3$ the independent standard deviations. Here, we excluded data above 9 $\mu m s^{-1}$ to obtain a higher fit quality of the significant measurements. The resulting peaks allowed us to divide the result into three populations: a significant fraction of particles showed speeds below 2$\mu m s^{-1}$ for all Au thicknesses, an intermediate fraction showed speeds in between 2 $\mu m s^{-1}$ and 6 $\mu m s^{-1}$ , and some of the particle coated with the 50 nm and 500 nm Au possessed speeds above 6 $\mu m s^{-1}$. This behavior differs quite a lot from what was observed at a lower fuel concentration and hints to an influence of the Au layer thickness on the particles' speed as well as at possible orientational differences. We will return to this in the next section.\\

Finally, from the distributions in Figure \ref{fig:fig3}B as well as from the sample trajectories in Figure \ref{fig:fig3}A, one can notice that at 5\%v/v H\textsubscript{2}O\textsubscript{2} some particles attain velocities higher than 9 $\mu m s^{-1}$ and reaching up to 15 $\mu m s^{-1}$. The speeds of these fast cuboids are comparable to those found in literature for magnetic and photo-activated haematite colloids with similar cap thicknesses\cite{Zhang2023}. We note that at a higher (8\%v/v) concentration of H\textsubscript{2}O\textsubscript{2}, the formation of bubbles interfered with data collection. 

\hspace{0.5 in}

\textbf{Sedimentation induced orientation and its effect on self-propulsion.} The results from the previous sections reveal a clear difference for our cuboid system between low and high fuel concentrations: at low fuel levels, the average particles' speed remains unchanged regardless of the increasing thickness of the Au layer, however, as the fuel level increases, two or three distinct populations emerge in the speed distribution, with two of them exhibiting significant speeds. These observations suggests that besides self-electrophoresis, other effects are dominating the particles' dynamics.

An anisotropic particle shape such as the cubic shape of our particles implies the possibility for different orientations of the particles and hence its director with respect to the surface. For particles made from materials with densities that are high compared to that of the surrounding solvent, thermal fluctuations can be insufficient to induce rotations of the particle after sedimentation. The orientation of the particle with respect to the surface during sedimentation then can become an important factor to consider, as particles with their director oriented out of the plan of the substrate will not be able to propel. We hypothesize that this is the origin of the different speed distributions seen at the different hydrogen peroxide concentrations. \\

The bimetallic coating induces a pronounced density difference throughout the particles' surface which affects their sedimentation. The silica cuboids not only are hollow, but Pt and Au are both on average 10 times more dense than silica. Differences in the thickness of the metal coating therefore are likely to influence the orientation of the cap with respect to the substrate during sedimentation. A similar torque has previously been observed for spheres with a pronounced mass anisotropy, which preferred to orient with the metallic cap towards the substrate when sedimenting in the absence of fuel\cite{Campbell2013,carrasco-fadanelli_sedimentation_2023,Sharan2022}. Indeed, Figures \ref{fig:fig2}A (A.3 and A.4) show the SEM micrograph of two different dried samples of bimetallic silica cuboids after left to sediment in water. Given the thickness and density of the metallic layers, the region corresponding to the silica base is pointing upwards, as signified by the dark circle on the center of the cubes. 

\begin{figure}[H]
    \centering
    \includegraphics[scale=1]{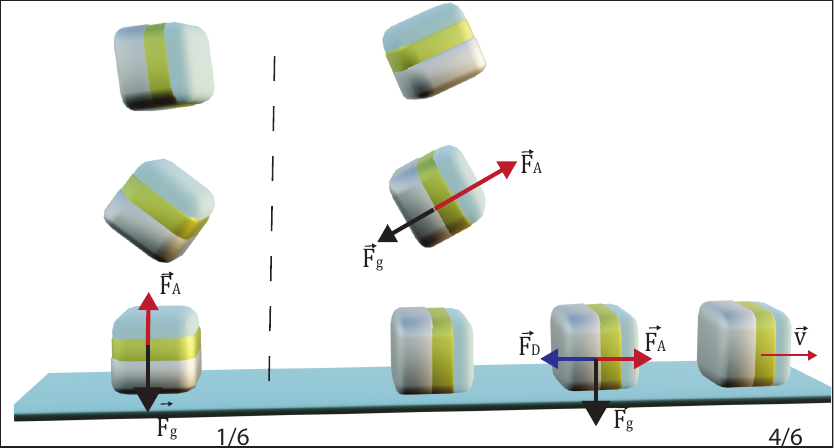}
    \caption{ Representation of the two possible sedimentation behaviors for bi-metallic silica-based cuboids. On the \textbf{left}, side the cuboids sediment with the metallic cap towards the substrate. This results in the active force ($\vec{F_{A}}$) pointing out of the plane, thus a low net displacement ($\vec{V}$). On the \textbf{right} side the cuboids sediment with the metallic cap perpendicular to the substrate, due to a stronger $\vec{F_{A}}$ which compensates for the gravitational torque ($\vec{F_{g}}$) while sedimenting, resulting in a particle orientation (4/6) that is favorable for in plane propulsion. The drag force ($\vec{F_{D}}$) is opposite to the active force.}
     \label{fig:fig4}
\end{figure}

The low velocities of the particles observed in the presence of 1\%v/v H\textsubscript{2}O\textsubscript{2} suggests that the mass anisotropy induces torques and thus cap-down orientations during sedimentation, even when the particle self-propel. Once the active cuboid particles have reached the substrate metal-side downward, their orientation with respect to the plane normal of the substrate plane must remain locked. Unlike many Janus particles which also show strong orientation quenching of the director with respect to the surface \cite{ketzetzi_diffusion-based_2020,ketzetzi_slip_2020,Das2015,Carreira2024}, the orientation of the cuboids is locked due to a combination of gravity and their anisotropic shape: once sedimented onto one side, rotation requires work against gravity. Their high density difference with respect to the solvent density leads to small gravitational heights and thus effective confinement. The active force $\vec{F_{A}}$ which points upwards out of the substrate's plane apparently is not able to induce rotations either, possibly due to hydrodynamic flows that usually lead to orientation quenching, see Figure \ref{fig:fig4}, left side. Low velocities could also stem from the scenario in which the $\vec{F_{A}}$ vector points into the plane, but are unlikely given the high mass anisotropy. \\ 

In contrast, at 5\%v/v H\textsubscript{2}O\textsubscript{2} there are two or three distinct speed populations of particles. The low-speed population, under 2$\mu m s^{-1}$, shows similar speeds as the particles at 1\%v/v H\textsubscript{2}O\textsubscript{2}, suggesting that their orientation on the substrate after sedimentation is similarly with their metallic cap down, and their active force vector is pointing out of the plane (left side of Figure \ref{fig:fig4}).  

The higher speeds of up to 9 $\mu m s^{-1}$ in the other populations on the other hand could correspond to cuboids with their active force vector being oriented along the plane. This implies that the gravitational torque either during sedimentation or after was counteracted and thus not able to reorient those particles. The higher active force at this higher fuel concentration is likely to be the origin of this as it can (partially) compensate the gravitational force during sedimentation.

This leads to a broader distribution of orientations during sedimentation resulting more often in particle orientation with the metal cap on the side on the substrate (Figure \ref{fig:fig4}). Such an orientation can harness the activity as is exemplified by the much larger speeds, which are on the order of two to four particle sizes per second, in line with ref. \cite{theurkauff_dynamic_2012, klongvessa_nonmonotonic_2019}. While different particle orientations can explain a slow and a fast population, it is still unclear why there are two distinct peaks in the speed distribution of the fast particles. Shape polydispersity inherent to the synthetic methods employed as well as slight differences of the metallic layers might explain these.\\
\hspace{0.5 in}

Thus, while two metals are needed for self-electrophoresis, a thickness of the base Au layer has only a minor effect on the self-propulsion speed. The overall thickness of the metals matters since it determines the mass anisotropy and thus gravitational torques. However, we find that the magnitude of the active force affect the sedimentation behavior of bimetallic silica-based cuboids and partially counteract the gravity induced torques. A higher activity allows for a higher fraction of cubes to land in orientations where the active director is parallel to the substrate and thus to propel faster, yielding the two behaviors observed.  

\begin{figure}[H]
    \centering
    \includegraphics[scale=1]{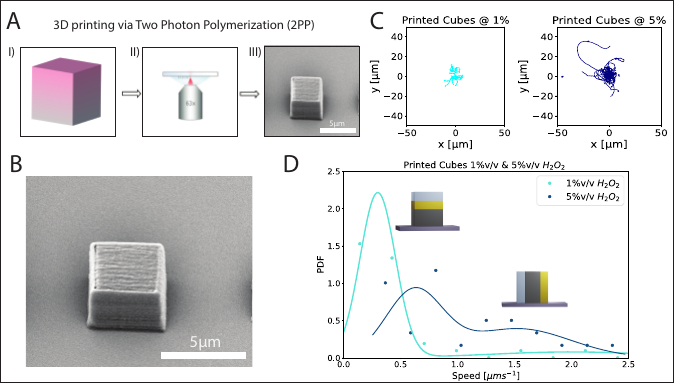}
    \caption{(A) Schematic of the printing procedure: (I) design of the Computer Assisted Design (CAD) model (II) printing using the Nanoscribe (III) development of the print to obtain the final shape. Adapted from \textit{Doherty et.al, 2020\cite{Doherty2020}}.  B) SEM micrograph of 3D printed cubes (4 $\mu m$ side length). (C) Trajectories centered at the origin, recorded over the course of 30 s at 20 fps, in the presence of 1\%v/v and 5\%v/v H\textsubscript{2}O\textsubscript{2}, for active printed cubes coated with a 10 nm Au base layer followed by a  10 nm Pt top layer. (D) Speed distributions of printed cubes coated with a 10 nm Au base layer followed by a  10 nm Pt top layer, in the presence of 1\%v/v and 5\%v/v H\textsubscript{2}O\textsubscript{2}. Speeds were obtained from fitting $MSD= 4D_{0}\delta t+ 2v^{2}\delta t^{2}$  with $\delta t_{max}=$ 1s, to the trajectories recorded via optical microscopy. We fit the data with a sum of two Gaussian distributions:  $f(x) = A_1 e^{-\frac{(x - \mu_1)^2}{2 \sigma_1^2}} + A_2 e^{-\frac{(x - \mu_2)^2}{2 \sigma_2^2}}$, where $A_1$ and $A_2$ are amplitudes, $\mu_1$ and $\mu_2$ the means and $\sigma_1$ and $\sigma_2$ the standard deviations.}
    \label{fig:fig5}
\end{figure}

\textbf{Dynamics of bimetallic Au-Pt 3D-printed cubes}. To test the influence of size and definition of the cubic shape as well as extend the technique to other materials, we apply the same bimetallic coating onto 3D printed particles with a cubic shape. 
3D printing via two-photon lithography allows tuning the shape parameter of the cuboid such that it becomes a perfect cube with sharp edges and corners ($m\gg 1$). 
We design cubes with a CAD program and print them using a Two-Photon-polymerization based mechanism (Nanoscribe Photonic Professional Gt), as depicted in Figure \ref{fig:fig5}A. The resulting particle is shown in Figure \ref{fig:fig5}B, and has a side length of 4 $\mu$m. 
\hspace{0.5 in}

 As before, we coated the cubes with a thin bimetallic layer consisting of a 10 nm Au layer at the base and 10 nm Pt layer on top. We observed the motion of these bimetallic 3D-printed cubes using brightfield microscopy  microscope at 1\%v/v and 5\%v/v H\textsubscript{2}O\textsubscript{2} and analyzed their trajectories using TrackPy. 
 Sample trajectories and speed distributions can be found in Figure \ref{fig:fig5}C and \ref{fig:fig5}D. For the active bimetalic 3D-printed cubes, we can observe  two-populations similar to what has as previously been seen for the active bimetallic silica-based cuboids under the same conditions: at  1\%v/v H\textsubscript{2}O\textsubscript{2} most of the cubes sediment in an orientation that yields no or low net displacement. However, at 5\%v/v H\textsubscript{2}O\textsubscript{2} the stronger active force again leads to faster particles, suggesting that it can (partially) balance the gravitational torque yielding more frequently an orientation that allows in plane self-propulsion. Therefore under similar conditions, there is no significant difference between the active mechanism for 3D-printed cubes with sharp edges and synthetic cuboids with rounded edges, despite the variation in size and material. For both fabrication techniques, self-propulsion is dominated by the sedimentation behavior, which leads to locked particle orientations on the substrate due to a combination of shape and gravitational confinement.\\
 
 The main difference between the 3D-printed system and the silica-based system is the overall lower speed: the peaks in the speed distribution for the 3D-printed cubes at a high fuel concentration are located at 0.62 $\mu m s^{-1}$ and 1.53 $\mu m s^{-1}$, respectively, and the average speed of the system is 1.03 $\pm$ 0.58 $\mu m s^{-1}$ which is considerably lower than for the synthetic cuboids (refer to Figure \ref{fig:fig3}C). We rationalize this lower value considering that the drag force experienced by a cubic particle is size dependent. In a uniform flow field, the drag force is defined as $\vec{F}_{drag_{cube}}= -\xi^{T}_{cube} \vec{v}$ where $\xi^{T}_{cube}$ is the translational friction coefficient which was found to scale with the side length $L$ as $ \xi^{T}_{cube}= 1.384 \cdot 6 \pi \eta  L $ in simulations performed by \textit{Okada et al.} \cite{Okada2019}. Assuming that the active force is the same and given that the side length of 3D printed cubes is $L_{3D}=4$ $\mu$m and $L_{Si}$= 2.0 $\pm$ 0.3 $\mu$m  for the silica cuboids, we expect their speeds to relate as $v_{3D}/v_{Si}=L_{Si}/L_{3D}=(2.0 \pm 0.3)/4= 0.50\pm 0.08$. This agrees very well with the ratio of the measured mean speeds of 1.03 $\pm$ 0.58 $\mu m s^{-1}$ (observed for the 3D printed cubes) and 2.24 $\pm$ 1.61 $\mu m s^{-1}$ (for 5nm Au coated cuboids), which corresponds to 0.46 $\pm$ 0.42. 
Thus, we can attribute the different speed mainly to the size dependent drag. Other effects, such as drag differences due to more rounded or sharp edges, intrinsic hydrodynamics of the active system that cannot be described using the active/drag force alone, effects originated from the presence of the surface such as slip, charge effects and roughness\cite{ketzetzi_slip_2020}, or differences in the extent of the interface between the two metals seem to play a minor role.

\section{Conclusions}

In this work we investigated the 2D dynamics of active bimetallic Au-Pt, silica-based cuboids and 3D-printed cubes. These electro-catalytic colloids are active through self diffusio- and electro-phoresis. Unlike the majority of the existing experimental systems, our cubes do not rely on magnetic or UV actuation to have a persistent self-propulsion mechanism and can be fabricated from materials such as silica and polymers. The colloids were studied via bright-field microscopy allowing us to qualitatively describe and quantify their dynamics for different thicknesses of the Au layer at different fuel concentrations.\\
For bimetallic silica-based cuboids at 1\%v/v H\textsubscript{2}O\textsubscript{2} only a small fraction of the cubes exhibited significant activity, while the majority showed little to no net displacement. We argued that this behavior is related to the metal-side downward orientation that the particles adopt during and after sedimentation on the substrate, given the thickness and density of the metallic layers, and the shape dependent quenching of the particles' orientation once they reach the substrate.\\ 
At higher fuel concentrations of 5\%v/v H\textsubscript{2}O\textsubscript{2}, two or three populations appeared in the speed distribution for all thicknesses probed: a population with a similar, low speed as for low fuel concentrations stemming from particles oriented with their metal cap down, and one or two faster populations of particles with their metal cap oriented to the side. We hypothesized that the higher active force compensates and counteracts the gravitational force during sedimentation, resulting in a reduced and orientation dependent torque, which increases the probability for a particle orientation that is favorable for in plane propulsion.\\

3D printed cubes were studied to understand the effect of material, shape and size on the dynamics of cubic colloidal systems. The mechanism for active bimetallic printed cubes proved to be similar as that of their silica analogs, where the in plane behavior is determined by the sedimentation behavior, rather than self-electrophoresis, which can be tuned with different activity levels. We attributed the lower self-propulsion speeds of the 3D printed particles compared to those of the silica-based cuboids to the increased drag due to their larger size.\\

Our experimental realization of bimetallic self-electrophoretic active cubes does not rely on magnetic or light activation for persistent self-propulsion, important for setups in which having a UV source is not possible or not desired. The bimetallic coating approach can be straightforwardly applied to other particle shapes. However, in line with earlier work on spheres, we find that metal coatings induce torques during sedimentation due to the inherent mass anisotropy. This, together with the anisotropic shape can lead to a significant fraction of particles with low or no net speed. An active particle design thus should take this into account, for example by circumventing shape-induced orientational locking by designing shapes that can rotate away from the mass-anisotropy induced orientation on the substrate. 
\hspace{0.5 in}

\begin{acknowledgement}
This work was supported by the Netherlands Organization for Scientific Research (NWO/OCW), as part of the Vidi scheme (grant nr. 193.069). The authors thank Rachel Dohetry for her help with particle synthesis and 3D printing, Alexandre Morin for useful discussions, and Yogesh Shelke for providing us with the silica cuboids.
\end{acknowledgement}
  
\bibliography{Manuscript_v8}

\end{document}